\documentclass[fleqn,usenatbib]{mnras}

\usepackage{newtxtext,newtxmath}
\usepackage[T1]{fontenc}
\usepackage{ae,aecompl}

\usepackage{graphicx}	
\usepackage{amsmath}	
\usepackage{amssymb}	

\title[Disc traced by Blue Stragglers]{A-type stars in the Canada-France Imaging Survey II.\\ Tracing the height of the disc at large distances with Blue Stragglers}

\author[G. F. Thomas et al.]{Guillaume F. Thomas$^{1}$\thanks{E-mail: guillaume.thomas@nrc-cnrc.gc.ca}, Chervin F. P. Laporte$^{2}$\thanks{$\!\!$ CITA National Fellow}, Alan W. McConnachie$^{1}$,
\newauthor Benoit Famaey$^{3}$, Rodrigo Ibata$^{3}$, Nicolas F. Martin$^{3,4}$, Else Starkenburg$^{5}$ 
\newauthor Raymond Carlberg$^{6}$, Khyati Malhan$^{3}$ and Kim Venn$^{2}$
\\
$^{1}$NRC Herzberg Astronomy and Astrophysics, 5071 West Saanich Road, Victoria, BC, V9E 2E7, Canada\\
$^{2}$Departement of Physics and Astronomy, University of Victoria, Victoria, BC, V8P 1A1, Canada\\
$^{3}$Universit\'e de Strasbourg, CNRS, Observatoire astronomique de Strasbourg, UMR 7550, F-67000 Strasbourg, France\\
$^{4}$Max-Planck-Institut f\"ur Astronomie, K\"onigstuhl 17, 69117 Heidelberg, Germany\\
$^{5}$Leibniz Institute for Astrophysics Potsdam (AIP), An der Sternwarte 16, D-14482 Potsdam, Germany\\
$^{6}$Departement of Astronomy and Astrophysics, University of Toronto, Toronto, ON M5S 3H4, Canada
}

\date{Accepted XXX. Received YYY; in original form ZZZ}

\pubyear{2018}

\begin{document}
\label{firstpage}
\pagerange{\pageref{firstpage}--\pageref{lastpage}}
\maketitle

\begin{abstract}
We present the kinematics of Blue Straggler (BS) stars identified in the Canada-France-Imaging-Survey (CFIS), covering 4000 deg$^2$ on the sky in the $u$-band. The BSs sample, characterised through CFIS and Pan-STARRS photometry, has been kinematically decomposed into putative halo and disc populations after cross-matching with Gaia astrometry and SDSS/SEGUE/LAMOST spectroscopy. This decomposition clearly reveals the strong flaring of the outer Milky Way disc. In particular, we show that we can detect this flaring up to a vertical height of $|Z| \simeq 8$ kpc at a Galactocentric distance of $R\sim $27 kpc. While some small level of flaring is expected for extended discs built up by radial migration, we demonstrate that the very strong flaring of the Milky Way disc that we observe is likely more consistent with it being dynamically heated by the repeated passage of the Sagittarius dwarf spheroidal galaxy through the midplane. 
\end{abstract}


\begin{figure*}
\centering
  \includegraphics[angle=0,  clip, viewport= 0 0 680 789, width=13cm]{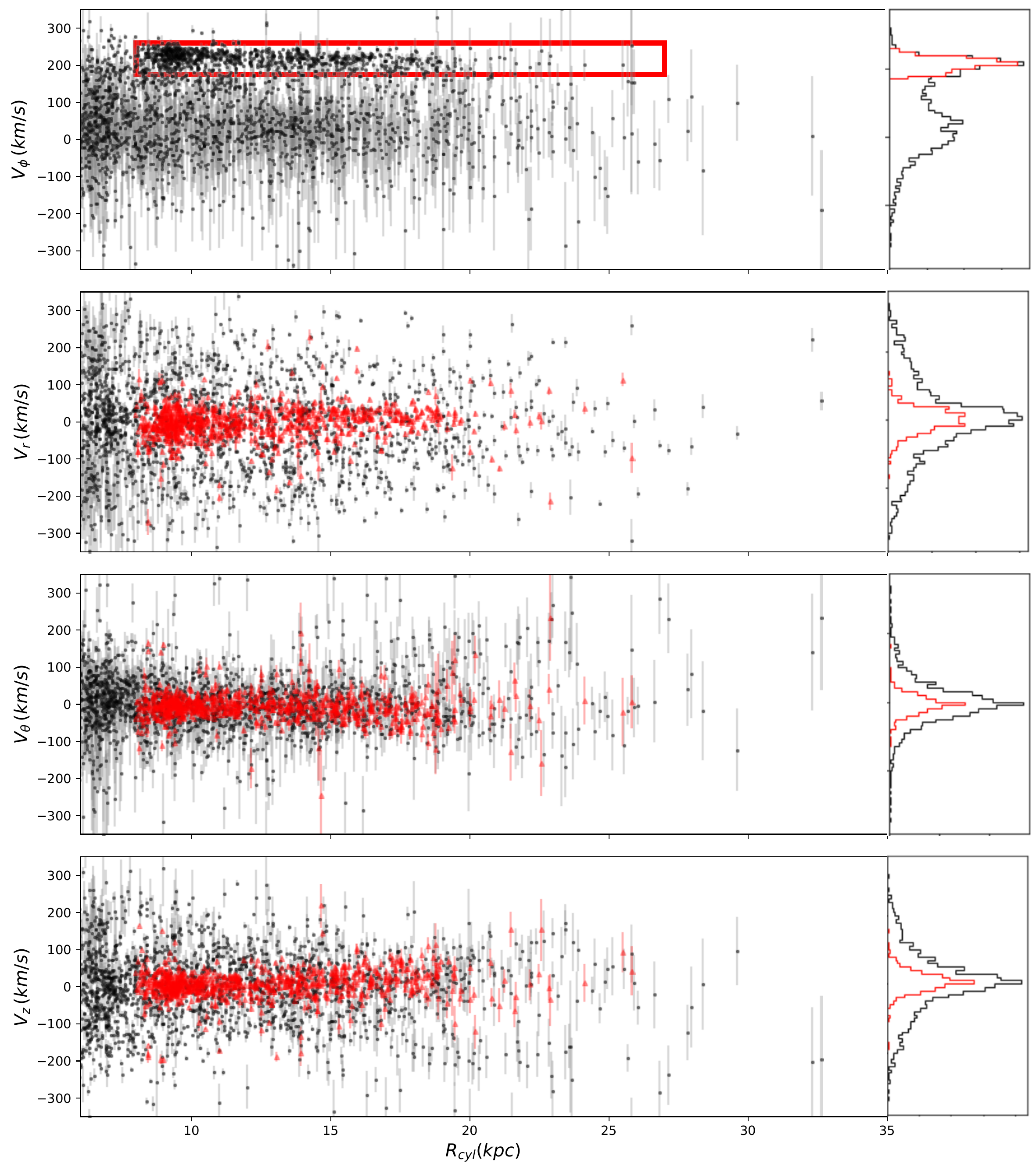}
   \caption{Galactocentric spherical and vertical (bottom panel) velocities of the 2258 BS stars with PMs and radial velocities, as a function of cylindrical radius in the plane of the disc. The red rectangle on the upper panel encompasses the candidate disc stars. These stars are shown in red in the three lower panels.}
\label{velocity}
\end{figure*}

\section{Introduction} \label{intro}

The Milky Way is generally considered to have formed, at least in part, by a succession of mergers. However, the details of its formation are still unclear and many questions remain. Among these, one unsettled problem is the fraction of stars in the halo of the Galaxy that formed {\it in situ}, and the fraction that originally came from the Galactic disc \citep[e.g.][]{eggen_1962,searle_1978,carollo_2007,carollo_2010,font_2011,rodriguez-gomez_2016,deason_2017a,pillepich_2018,belokurov_2018}. The stellar populations that are often used to trace the disc (e.g. Main Sequence Turn-off, MSTO) and the halo (e.g. Blue Horizontal Branch -- BHB, or K-giants) are quite different in terms of their evolutionary state. The former are very common and relatively faint, and are ideal for studying the details of our close environment, which is largely dominated by disc stars; the latter are relatively rare (they are only present in old, metal-poor stellar populations) and very bright, so that they can probe the halo out to very large distances. Using different stellar populations to characterise the disc and the halo lead to complications when attempting to understand the properties of structures at the interface of these two components. In this respect, it is desirable to use a stellar population which is both bright and present in relatively large numbers. 

The Blue Stragglers (BS) were observed for the first time by \citet{sandage_1953} and are A-type stars that appear bluer and brighter than MSTO stars of similar metallicity and age, but which have a surface gravity typical of dwarfs. Many mechanisms have been proposed to explain the formation of these anomalous bright stars, such as a merger of stars due to collisions in dense stellar systems \citep{hills_1976} or mass transfer between two or three stars in a multiple stellar system \citep{mccrea_1964}, but a consensus has not yet been reached \citep[see, e.g.,][ and references therein]{perets_2015}. They are more numerous and closer than typical halo BHB stars, hence they can have a similar apparent magnitude while being fainter by $\sim 2$~mag in absolute terms. They are often identified as a source of contamination in studies of other stellar types \citep[e.g.,][]{clewley_2004,bell_2010,deason_2011,deason_2014,fukushima_2018}, but since they are bright and present in all stellar components of the Galaxy, they are potentially ideal tracers of the disc/halo interface.

The first evidence of the flaring of the Galactic disc was discovered by \citet{lozinskaya_1963} who observed that the scale height of the atomic HI gas distribution of the disc increases with Galactic radius, reaching a scale height of 2.7 kpc at $\sim$ 40 kpc \citep{kalberla_2007}. Nevertheless, the amplitude of the flaring of the stellar component is still debated \citep{reyle_2009,polido_2013,kalberla_2014,lopez-corredoira_2014,amores_2017}. \citet{minchev_2012} showed that the Galactic disc flare can be a consequence of the radial migration of stellar populations in the thin disc component. However \citet{vera-ciro_2014} and \citet{vera-ciro_2016} argued that the inward migrators instead induce a thinning of the disk, rather than the outward migrators causing a thickening. In any case, the amplitude of the flaring due to radial migration is expected to be less pronounced than in the case of a flare caused by disc-satellite impacts \citep{ibata_1998,velazquez_1999, kazantzidis_2008,villalobos_2008,bournaud_2009,purcell_2010,gomez_2013}.

In this paper we study the kinematics of the disc and halo BSs populations identified with the Canada-France-Imaging-Survey (CFIS). We show that we can detect the flaring of the stellar component of the disc up to vertical heights of $|Z| \simeq 8$ kpc at a Galactocentric distance of $\sim$ 27 kpc. We suggest that this very strong flaring may be consistent with the dynamical heating of the disc caused by the orbit of the Sagittarius dwarf spheroidal galaxy (Sgr dSph) around the Galaxy. Such a process was initially proposed by \citet{ibata_1998}, and we compare our data to recent models of this process by \citet{laporte_2018b}.

\section{Data} \label{data}

The BSs sample used in this paper was identified photometrically by \citet{thomas_2018a} (hereafter, Paper I) using the $u$-band of the Canada-France-Imaging Survey (CFIS) \citep{ibata_2017a} and the \textit{griz} bands from Pan-STARRS 1 \citep{chambers_2016}. This photometric sample has been demonstrated to have low ($\sim 15$ \%) contamination from BHB stars.

In order to obtain proper motions (PMs) of the BSs catalog, a cross-match with the second data release of the Gaia mission \citep{gaiacollaboration_2018} is performed. Another cross-match with the SDSS/SEGUE \citep{yanny_2009a} and LAMOST (DR3) surveys to obtain line-of-sight velocities and metallicities is made. Our final sample of BSs with PMs, radial velocities and metallicities consists of $2612$ stars over the current CFIS footprint of $\simeq 4,000$ deg$^2$.

The CFIS footprint is shown on Figure 1 of Paper I. Note that it is not symmetric with respect to the Galactic plane, having more stars in the Northern Galactic hemisphere. After cross-matching with SEGUE and LAMOST, this North-South asymmetry becomes even stronger. Hence any North-South asymmetry in our results is potentially a consequence of the footprints of these surveys. 

\begin{figure}
\centering
  \includegraphics[angle=0,  clip, width=8cm]{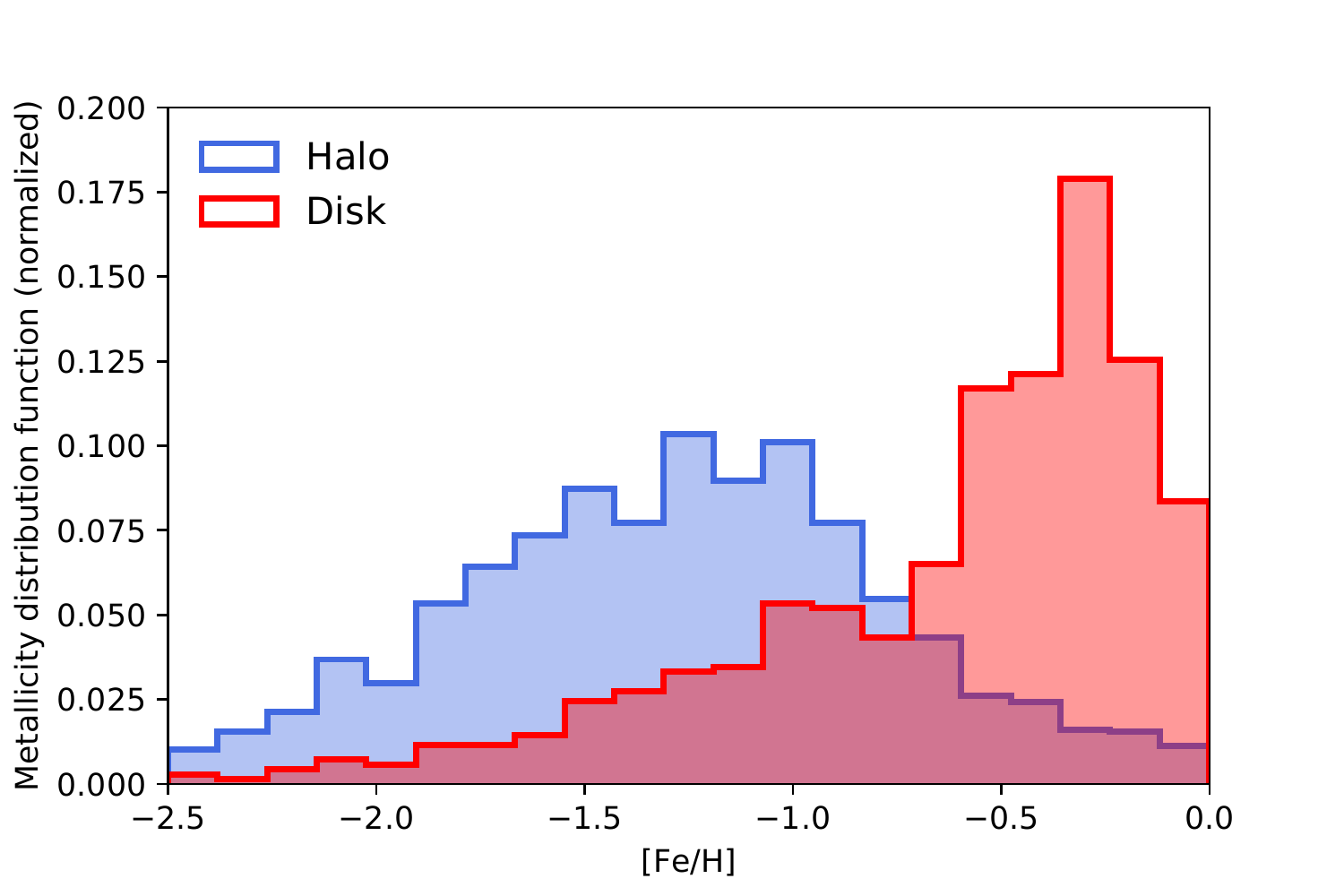}
   \caption{Normalized metallicity distribution functions of the putative disc BSs (red) and of the putative halo BSs (blue). The metallicities used here are the {\it FeHadop} from SDSS/SEGUE and the {\it feh} from LAMOST.}
\label{MDF}
\end{figure}

\begin{figure*}
\centering
  \includegraphics[angle=0,  clip, width=12cm]{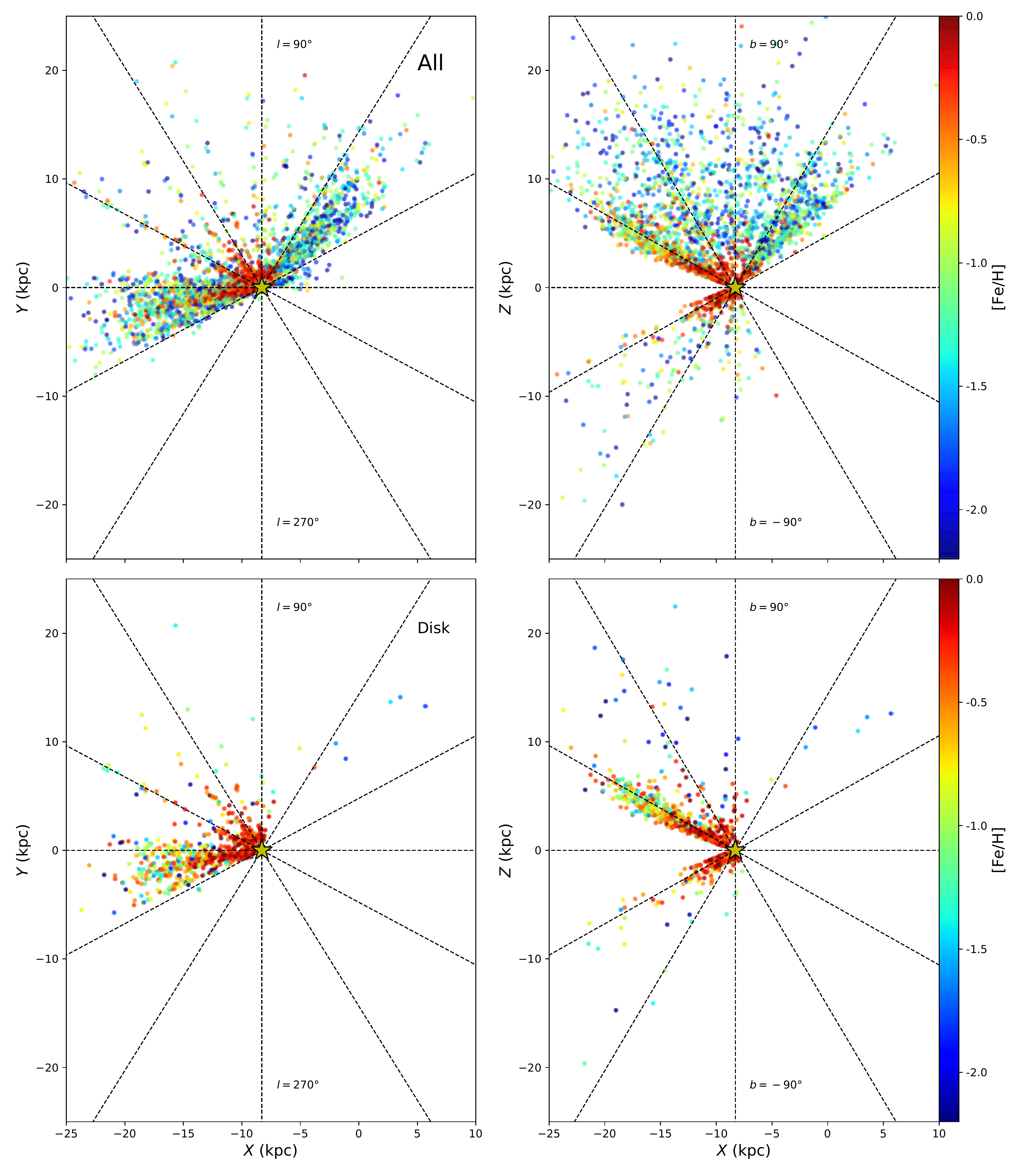}
   \caption{Projections of the Galactocentric distribution of all the BSs stars (top panels), and of the putative disc BSs (bottom panels), colour-coded by metallicity. The yellow star symbol shows the position of the Sun. The putative disc BSs stars, which are selected on a purely kinematic basis, are closer to the plane than the rest of the BSs and are more metal-rich than the halo BSs, as expected for a disc population.}
\label{position}
\end{figure*}
$ $
\begin{figure*}
\centering
  \includegraphics[angle=0,  clip, viewport= 0 82 1020 645, width=14cm]{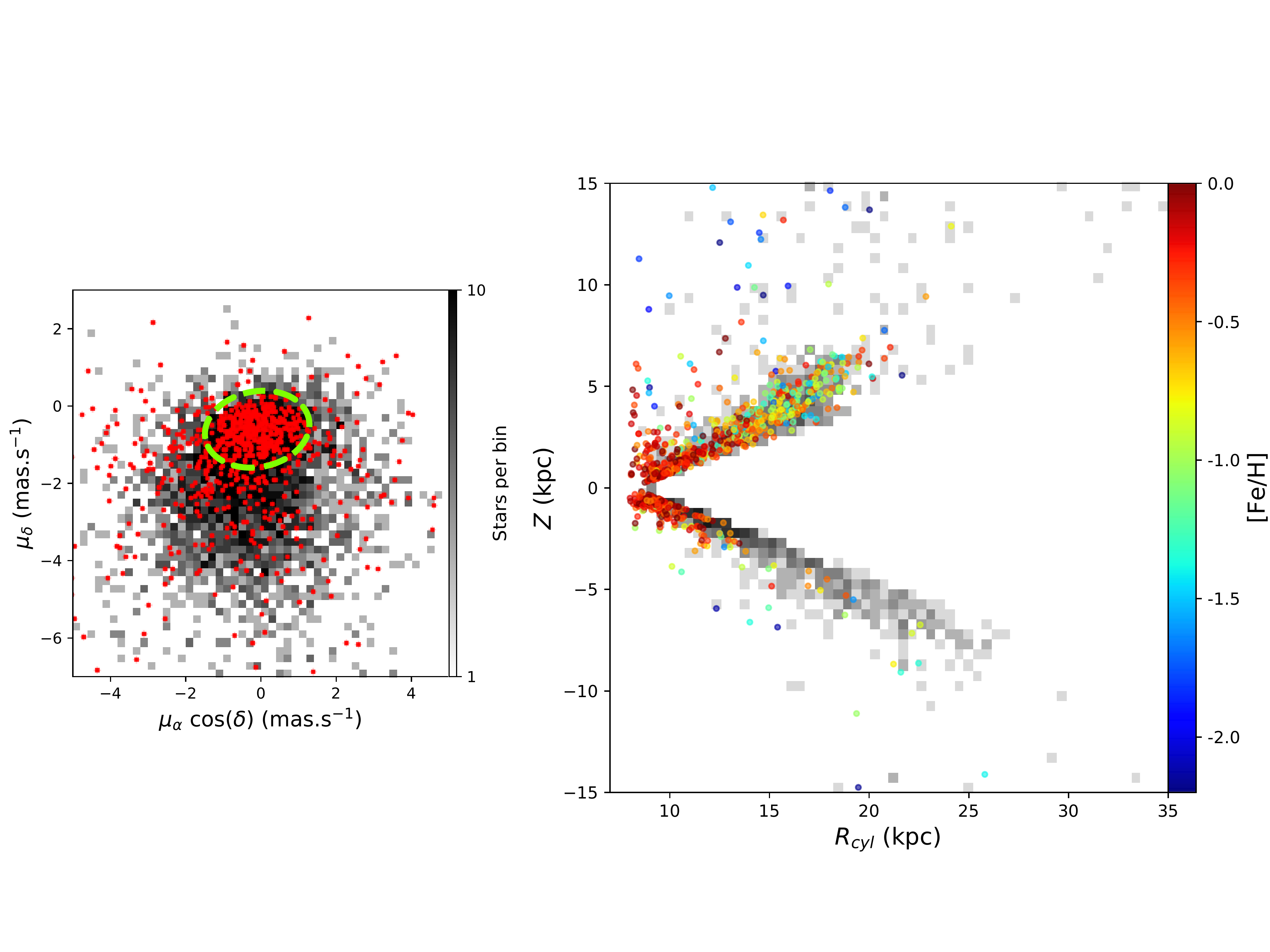}
   \caption{{\bf Left panel:} proper motions of the BSs from CFIS located between $140 \degr \leq l \leq 240 \degr$ in gray and proper motion of the identified disc BSs from the spectroscopic sample in red. The green ellipse show the selection of potential disc BSs selected only by their proper motion. {\bf Right panel:} $z -- R$ projection of disc BSs, identified by their PM (in grey) and by their circular velocity (coloured points).}
\label{PM}
\end{figure*}
$ $
\begin{figure}
\centering
  \includegraphics[angle=0,  clip, width=8cm]{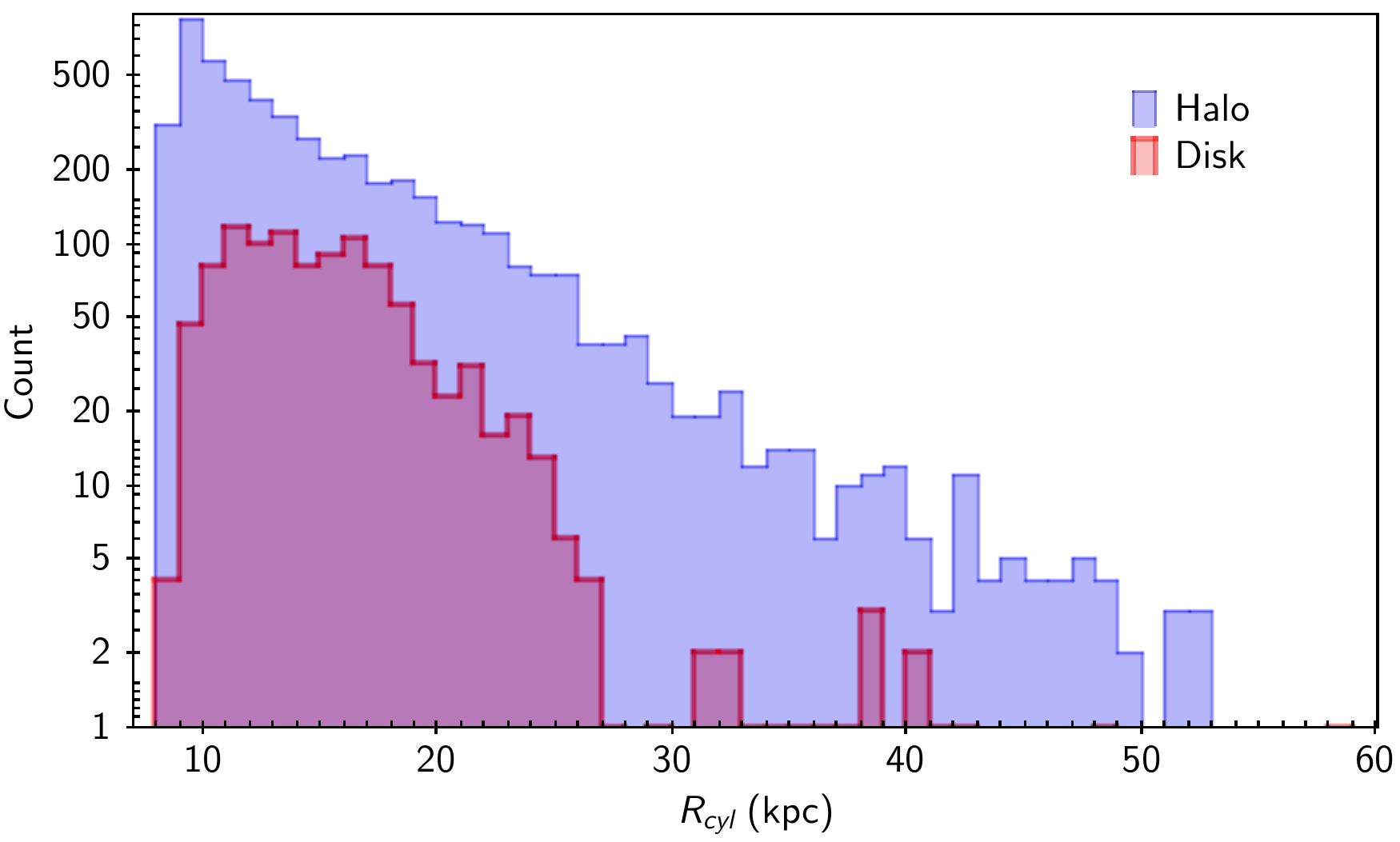}
   \caption{Galactocentric distance distribution of the disc BSs identified by their proper motion (in red) compared to the remaining BSs population (in blue) which we consider to be the halo population.}
\label{dist_PM}
\end{figure}
$ $

\section{Method and results} \label{method}

\subsection{Distances estimates}

To derive tangential velocities from the PMs of our stars, it is necessary to determine their distance. For faint sources the Gaia DR2 parallax measurements have large uncertainties. For a typical BSs with $(g-r)_{0,SDSS}=-0.1$ and $M_g=2.56$ at 10 kpc, the parallax uncertainties are of the same order as the measurements themselves (expected Gaia end-of-mission accuracy)\footnote{We have used the color equations of \citet{jordi_2010} to transform from the SDSS $g, r$ to Gaia $G$}. Therefore, photometric distances have to be determined instead. This is done using the calibration of the absolute magnitude of the BSs in the $g$-band provided by \citet{deason_2011}, using the Sagittarius stream lying in the SDSS Stripe 82:
\begin{equation}
M_g = 3.108+5.495 \, (g-r)_{0,SDSS} \, .
\end{equation} 
As discussed in Paper I, we recalibrate the $(g-r)_{0}$ colour from the Pan-STARRS 1 photometric system to that of the SDSS.

The accuracy of our distances is accessed by comparing them to previous distance estimates for two globular clusters, NGC 5272 and NGC 5466. Both these clusters are present in the CFIS footprint and a population of BSs can be identified in each of them (see figure 4 of Paper I). For NGC 5272, we find a mean BSs distance of $10.4$ kpc, in agreement with the distance estimated by \citet{harris_1996} of $10.2\pm 0.2$ kpc. For NGC 5466, we find a distance of $16.8$ kpc also consistent with the value of $16 \pm 0.4$ kpc listed by \citet{harris_1996}. The dispersion of our estimated individual distances of BSs in these two clusters is 1.9 kpc and 2.3 kpc, respectively. Based on these values, we estimate that the BS-based distances have a relative precision of approximately $20$\%.

\subsection{Selection of disc stars}

The Galactic Cartesian coordinates ($X$, $Y$, $Z$) and velocities ($Vx$, $Vy$, $Vz$) are defined using the conventions adopted in the \textsf{astropy} package \citep{theastropycollaboration_2018}, with the most recent estimate of the Sun's position ($X = -8.1$ kpc, \citealt{gravitycollaboration_2018a}) and with a circular velocity at that position of $229.0$ km.s$^{-1}$ \citep{eilers_2018}. The adopted Solar peculiar motion is that of \citet{schonrich_2010}, namely ($U_\odot$, $V_\odot$, $W_\odot$) = (11.1, 12.24, 7.25) km.s$^{-1}$ in Local Standard of Rest coordinates.

The spherical Galactocentric coordinates (r, $\theta$, $\phi$)\footnote{Here $\theta$ is the elevation angle and $\phi$ is the azimuthal angle.} and velocities ($V_r$, $V_\theta$, $V_\phi$) for our sample are determined in the same way as \citet{bird_2018}. We propagate the uncertainties in PM, line-of-sight velocities and distances by calculating the mean dispersion in the Galactic positions and velocities from $1000$ Monte-Carlo realisations, selecting from Gaussian distributions of each of the original quantities.

Figure \ref{velocity} shows the 3D Galactic velocities as a function of the cylindrical distance in the plane of the disc ($R_{cyl}= \sqrt{X^2 + Y^2}$). Two components are clearly visible on the bimodal histogram of the upper panel. The first one is centred at $V_{\phi}=0$ km.s$^{-1}$ with a large dispersion, and likely corresponds to the BSs of the stellar halo. The second component, highlighted by the red rectangle, has a mean rotational velocity of $V_{\phi}\simeq 220$ km.s$^{-1}$ and a relatively small velocity dispersion compared to the putative halo population. It appears, therefore, that this population is {\it a priori} consistent with the disc, although it extends to large radii ($\gtrsim 25$kpc). Of course, such a pure kinematic selection must also contain some halo BSs. It is therefore interesting to check for the spatial and metallicity distribution of this kinematically-selected disc-like population.

The normalised metallicity distribution functions (MDFs) of the putative disc BSs and the putative halo BSs are shown in Figure \ref{MDF}. The MDF of the putative disc is close to the MDF of the populations of the disc identified by \citet{ibata_2017b} (populations $A$ and $B$ in their terminology) and the MDF of the putative halo is similar to their halo MDF (population $C$). This thus strengthens the interpretation of these components as being disc-like and halo-like. 36 BSs are in common in the SDSS/SEGUE and LAMOST catalogs in the CFIS footprint. The spectroscopic metallicities given in these two catalogs have a difference of typically $\Delta[Fe/H]=$0.25, in the $1-\sigma$ of the LAMOST uncertainties. Nevertheless, the difference in the metallicity between the two catalogs does not impact our interpretation, since the scale of this difference in metallicity between the disc-like and halo-like components being larger than the difference between the two spectroscopic catalogs. 

Figure \ref{position} shows the spatial distribution of our BSs populations in Galactic Cartesian coordinates, colour-coded by their metallicity. The two top panels show the whole BSs sample, whilst the bottom panels show the kinematically-selected disc population. The kinematically detected disc BSs stars are located at smaller vertical distances (typically $|Z|< 8.5$ kpc) than the halo BSs, which are more homogeneously distributed. The fact that a small proportion of the kinematically-selected disc is composed of low-metallicity stars at large $Z$ is not very surprising, since one should expect some contamination from the halo component even for a sample selected with large $V_{\phi}$.

As mentioned by \citet{xue_2008}, a fraction of photometrically identified BSs can be, in fact, young Main Sequence stars. However, for a typical population of the disc with [Fe/H]=-0.5, a hot MS star of age 1 Gyr has an absolute magnitude of $M_g=1.5$\footnote{In the SDSS photometric system.} and $M_g=2.2$ for a MS star of 2 Gyr, similar to or brighter than the disc BSs which have an absolute magnitude in the range $M_g= 3.8-2.2$. Therefore the estimated distance of these contaminating young MS stars will be smaller than what they really are, making the flare discussed hereafter even stronger. For older populations, the hotter MS stars are too cold to be confused with BS stars.

The bottom panels of Figure \ref{position} show that the disc flares toward the Galactic Anticentre and reaches a height of $Z \simeq 6.5$ kpc at a Galactocentric distance of 23 kpc in the plane of the disc. \citet{juric_2008} and \citet{ivezic_2008}, using MS stars, have observed a similar flaring of the disc beyond $R>14$ kpc and associated this flaring to the Monoceros Ring (also known as the Galactic Anticenter Stellar Structure, or GASS, \citealt{crane_2003}). \citet{li_2017} show that the Monoceros Ring has similar properties to A13 \citep{sharma_2010} and the Triangulum-Andromeda overdensity \citep[TriAnd,][]{martin_2007}. Together, they form a unique structure generated by the oscillation of the disc due to the passage of the Sgr dSph through it \citep{gomez_2016,laporte_2018,laporte_2018b}. A disc origin for this structure is consistent with the measured fraction of RR Lyrae stars to M giants  \citep{sheffield_2018}, which would otherwise be difficult to reconcile with a senario where the structure formed from the tidal disruption of a dwarf galaxy\citep[e.g.][]{penarrubia_2005}. \citet{li_2017} found that the M-giant stars in the Moneceros-A13-TriAnd structure have a low velocity dispersion ($< 40$ km.s$^{-1}$) and display a gradient in their radial velocity as a function of Galactic longitude. Although a small number ($\sim$ 10) of our disc BSs show a velocity gradient at the distance of the Monoceros Ring \citep{crane_2003}, the majority of them does not follow this trend. Rather, they display standard disc-like kinematics, consistent with a classical flare.

Most of the disc BSs identified are located towards the Galactic Anticentre. In this direction, the rotation velocity is dominated by the PM. In order to check how far the disc component might extend in this direction, it is possible to select a larger sample of disc BSs using only their PM. This selection will be more complete than the spectroscopic sample, since it will not limited by the Segue/LAMOST selection function and will not biased towards the Galactic north as the previous spectroscopic sample.

The greyscale in the left panel of Figure \ref{PM} shows the PM distribution of all $1397$ BSs from the CFIS+Gaia sample located between $140 \degr \leq l \leq 240 \degr$. The red points show the PMs of the disc BSs selected in Figure 1. The dashed circle shows a selection contour inside of which disc BSs are preferentially selected, without the need for line-of-sight velocities. 

\begin{figure*}
\centering
\includegraphics[angle=0, clip, viewport= 0 0 1020 556, width=14.5cm]{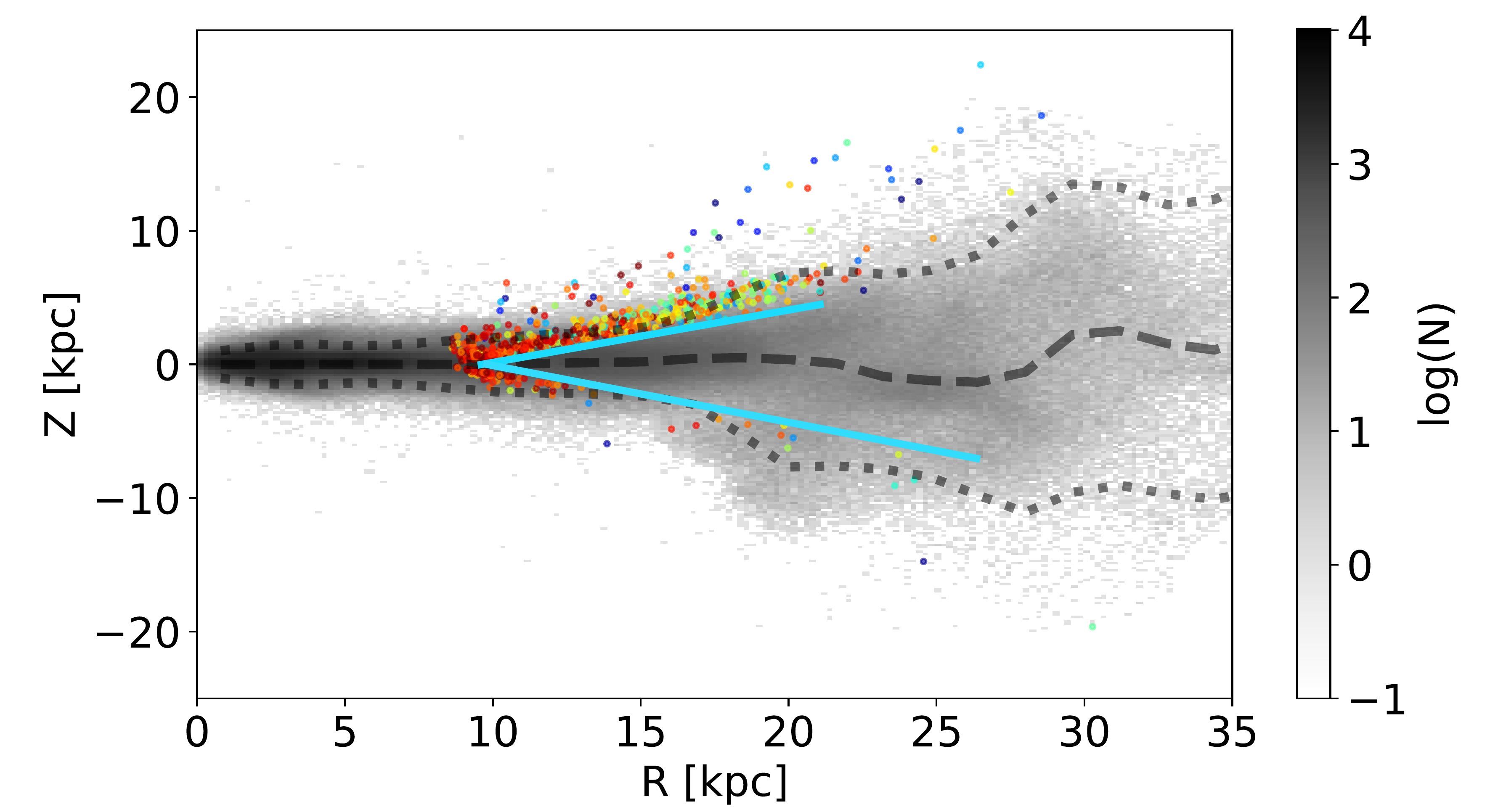}
\caption{The projected $Z-R$ distribution of a Galactic disc perturbed by a Sgr-like dwarf is shown in grey from \citet{laporte_2018b}. The median height of the disc and $1$-$\sigma_Z$ dispersion as a function of Galactocentric radius are shown by the dashed and dotted lines, respectively. The two thick blue lines show the limit of the CFIS footprint toward the Galactic Anticentre and their length correspond to the the maximum length of the disc detected in the PM-selected disc BSs sample (see Figure \ref{PM}). The disc BSs are colour-coded by metallicity with the same scale as in Figure \ref{PM} .}
\label{sgr_pos}
\end{figure*}

The greyscale in the right panel of Figure \ref{PM} shows the $Z - R$ distribution of the BSs selected via the PM cut in the left panel. The points show the $Z - R$ distribution of the previous disc BSs shown in Figure 3. As a result of the PM-only selection, we see the disc in the southern hemisphere also exhibits the stellar disc flare identified previously. We thus demonstrate that, although wobbles in the background, such as Monoceros, do exist, the general strong flaring of the disc is present in both hemispheres. Analyses based on SDSS alone were not able to probe beyond Galactocentric radii $R \geq 12$ kpc in the South \citep{juric_2008}.

It is worth noting here that the flaring detected in both hemispheres in Figure \ref{PM} can be, in fact, the signature of the warping of the disc at different Galactic longitudes. Indeed, since the CFIS footprint covers the southern Galactic hemisphere between $140\degr \leq l \leq 170 \degr$ and in the northern hemisphere between $170\degr \leq l \leq 240 \degr$, the warping of the disk can generate a similar signature than seen on Figure \ref{PM}. However, \citet{momany_2006} measure the maximum (absolute) warp height to be 0.6 kpc at $l \simeq 240 \degr$ at a distance of 16 kpc from the Sun. This is inconsistent with the height of the BSs as seen with our data in both hemispheres.

The PM-only selection is not limited in magnitude in the same way as the spectroscopic selection is. Indeed, with CFIS and Gaia, BSs can be identified up to a distance of 53 kpc. However, we see clearly in Figure \ref{dist_PM} that the number of BSs that have PM consistent with the disc fall rapidly beyond a distance of $\simeq 27$ kpc from the Galactic centre, i.e. it looks as if we are sampling BSs out to the furthest ``edge'' of the disc. Nevertheless, due to the current CFIS footprint, it is not possible to determine the radial scale length of the vertical fluctuation, or the vertical scale height as a function of the radius. This may need to wait future extension of the CFIS footprint or other surveys. 

\section{Discussion and Conclusions}

While some flaring is expected in the disc outskirts for extended discs built up by radial migration \citep{minchev_2012,minchev_2012a}, the scale of this effect is in no way comparable to the amount of flaring that we find in the Milky Way outskirts. In this section, we interpret our findings as possibly being related to the dynamical heating of the disc by a passing satellite galaxy such as the Sagittarius dwarf.  

Hereafter, we compare our results with one of the N-body simulations taken from the suite of \citet{laporte_2018b}. These simulations model the interaction of the Milky Way with different initial progenitors of a Sagittarius-like dSph, for which the remnants have a central velocity dispersion consistent with $\sigma \lesssim 20$ km.s$^{-1}$. These were also complemented with two additional runs taking into account the contribution of the Large Magellanic Cloud, assuming a first infall orbit \citep{besla_2007, kallivayalil_2013}. In these models, the Sgr dSph produces vertical oscillations as well as coupled radial motions in the disc \citep{donghia_2016} through the response of the MW halo \citep{weinberg_1998, vesperini_2000, gomez_2016} to its infall and eventually through its tidal interactions.

These models were the first ones to successfully reproduce the outer disc structures, such as the spatial extents of the Monoceros Ring, A13 and TriAnd \footnote{As mentioned in the previous section, these have recently been confirmed to have stellar populations, kinematics and abundance measurements consistent with a disc population \citep{price-whelan_2015a, li_2017, sheffield_2018, bergemann_2018,deboer_2018,deason_2018}, while satisfying the constraints set by kinematics and amplitudes of the spatial number density counts in the solar neighborhood (\citealt{widrow_2012}, but see also \citealt{schonrich_2018}). Thus, they established a link, and a possible common origin, between the outer disc structures and the local vertical disc oscillations, as suggested in earlier works \citep{purcell_2011, gomez_2013,gomez_2016}.} They also produced structures reminiscent of the Anticentre Stream \citep[ACS,][]{grillmair_2006b} interpreted as remnants of tidal tails \citep{laporte_2018b}. Shortly after the second Gaia release, \citet{laporte_2018a} also showed that the last pericentric passage of Sgr with the MW could seed the disc perturbations that give rise to the observed velocity field in the Gaia Volume \citep{gaiacollaboration_2018a}, as well as the formation of the phase-space spiral uncovered by \citet{antoja_2018}, finding similar timescales for its onset\footnote{Clues for incomplete phase-mixing in the Milky Way and its relation with Sgr were already discussed in the context of the UV plane \citep[e.g.][]{minchev_2009,gomez_2012}.}.

In Figure \ref{sgr_pos}, the distribution of our BSs sample is compared to one of these simulations in the $R-Z$ plane for stars selected within a wedge of $140^{\circ}<l<240^{\circ}$ and which have a kinematic signature typical of the disc. Here, we reproduce the model in greyscale, while the coloured points show the observed data. Dashed lines mark the ``midplane'' of the model disc (median height of the stellar plane) together with the corresponding $1\sigma$ standard deviation curves. The BSs disc stars appear to follow the flared structure of the model of the outer disc, including some of the most distant stars in our sample.

In summary, we have disentangled the halo and disc BSs populations observed in CFIS, using Gaia, SDSS/Segue, and Lamost data. This allows us to trace the disc to large radius and we find that the disc height at a Galactocentric radial distance of $\sim 27$ kpc reaches up to $|Z|\sim 8$ kpc above the midplane. By comparing these observations to the simulations of \citet{laporte_2018b}, we have shown that this flare of the disc could potentially be explained via dynamical heating of the disc by successive passages of the Sagittarius dwarf galaxy about the midplane of the disc, consistent with early studies of disc-satellite impacts \citep[e.g.,][]{ibata_1998,velazquez_1999, kazantzidis_2008,villalobos_2008, purcell_2010} and expectations from cosmological simulations \citep[e.g.,][]{aumer_2013,gomez_2016}.

\section*{Acknowledgements}
We thank the anonymous referee for their careful reading and for their helpful and constructive comments.\\
We also thank Julio Navarro for his interesting suggestions and comments. 
\\
This work used the Extreme Science and Engineering Discovery Environment (XSEDE), which is supported by National Science Foundation grant number OCI-1053575.\\
This work is based on data obtained as part of the Canada-France Imaging Survey, a CFHT large program of the National Research Council of Canada and the French Centre National de la Recherche Scientifique. Based on observations obtained with MegaPrime/MegaCam, a joint project of CFHT and CEA Saclay, at the Canada-France-Hawaii Telescope (CFHT) which is operated by the National Research Council (NRC) of Canada, the Institut National des Science de l'Univers (INSU) of the Centre National de la Recherche Scientifique (CNRS) of France, and the University of Hawaii, and on data from the European Space Agency (ESA) mission {\it Gaia} (\url{https://www.cosmos.esa.int/gaia}), processed by the {\it Gaia} Data Processing and Analysis Consortium (DPAC,
\url{https://www.cosmos.esa.int/web/gaia/dpac/consortium}). Funding for the DPAC has been provided by national institutions, in particular the institutions participating in the {\it Gaia} Multilateral Agreement.\\We also used the data provide by the Guoshoujing Telescope (the Large Sky Area Multi-Object Fiber Spectroscopic Telescope LAMOST), a National Major Scientific Project built by the Chinese Academy of Sciences. Funding for the project has been provided by the National Development and Reform Commission. LAMOST is operated and managed by the National Astronomical Observatories, Chinese Academy of Sciences.\\
Funding for SDSS-III has been provided by the Alfred P. Sloan Foundation, the Participating Institutions, the National Science Foundation, and the U.S. Department of Energy Office of Science. The SDSS-III web site is http://www.sdss3.org/.
SDSS-III is managed by the Astrophysical Research Consortium for the Participating Institutions of the SDSS-III Collaboration including the University of Arizona, the Brazilian Participation Group, Brookhaven National Laboratory, Carnegie Mellon University, University of Florida, the French Participation Group, the German Participation Group, Harvard University, the Instituto de Astrofisica de Canarias, the Michigan State/Notre Dame/JINA Participation Group, Johns Hopkins University, Lawrence Berkeley National Laboratory, Max Planck Institute for Astrophysics, Max Planck Institute for Extraterrestrial Physics, New Mexico State University, New York University, Ohio State University, Pennsylvania State University, University of Portsmouth, Princeton University, the Spanish Participation Group, University of Tokyo, University of Utah, Vanderbilt University, University of Virginia, University of Washington, and Yale University. 
This research used the facilities of the Canadian Astronomy Data Centre operated by the National Research Council of Canada with the support of the Canadian Space Agency.

R. A. Ibata and N. F. Martin acknowledge support by the Programme National Cosmology et Galaxies (PNCG) of CNRS/INSU with INP and IN2P3, co-funded by CEA and CNES. This work has been published under the framework of the IdEx Unistra and benefits from a funding from the state managed by the French National Research Agency as part of the investments for the future program.\\ E. S. gratefully acknowledges funding by the Emmy Noether program from the
Deutsche Forschungsgemeinschaft (DFG).

\label{lastpage}

\bibliography{./biblio}

\end{document}